\documentclass{jpsj-suppl}
\usepackage{txfonts} 

\newcommand{\kslash}{k \hspace{-5pt} / \hspace{2pt}}

\title{Mesons in Nuclei and Partial Restoration of Chiral Symmetry}

\author{Daisuke \textsc{Jido}}

\inst{Department of Physics, Tokyo Metropolitan University, Hachioji, Tokyo 192-0397, Japan}

\email{jido@tmu.ac.jp}

\recdate{January 23, 2016}

\abst{Recent topics on mesons in nuclei are discussed by especially emphasizing the role of the partial restoration of chiral symmetry in the nuclear medium. The spontaneously broken chiral symmetry in vacuum is considered to be incompletely restored in finite nuclear density systems with moderate reduction of the magnitude of the quark condensate. On the partial restoration of chiral symmetry, the wave function renormalization is important to be taken into account for the Nambu-Goldstone bosons. We also discuss the possible change of the meson properties in the nuclear medium and meson-nucleus systems for the $\bar K$, $\eta$, $K^{+}$ and $\eta^{\prime}$ mesons.}

\kword{chiral symmetry, partial restoration, nuclear medium, hadrons in nuclei, $\eta$~meson, $\bar K$ meson, $K$ meson, $\eta^{\prime}$meson}

\begin{document}
\maketitle

\section{Introduction}

The study of the hadron properties in nucleus attracts us in many different aspects. 
Nuclear physics investigates many-body systems governed by the strong force
and aims at understanding the nature of the strong interaction. 
With this knowledge, hopefully, new bound systems of the strong interaction are desired 
to be discovered and predicted. Such exotic states themselves have attracted our continuous attentions. 
Hadrons in a nucleus change their properties due to the strong interaction with nucleons. 
These in-medium modifications can be described 
by hadronic many-body effective theories, since hadrons are fundamental degrees of the system in low energies
thanks to the color confinement. Nevertheless, in hadron physics it is extremely important to have its more
fundamental interpretation in terms of QCD. Its key concepts are spontaneous breaking of chiral symmetry 
and its partial restoration in the nuclear medium.
The study of hadrons in nuclei also provides basic informations of high density physics and 
gives important constraints accessible in our reaction experiments. 

Chiral symmetry is one of the fundamental symmetries in the underlying theory of the strong interaction, QCD, 
and is considered to be dynamically broken by the physical states. 
The dynamical chiral symmetry breaking determines vacuum properties and describes low-energy hadron dynamics. 
The dynamical breaking of chiral symmetry is a phase transition phenomenon and 
it can be restored at extreme conditions, such as high density and high temperature. 
The broken chiral symmetry is also considered to be partially, or incompletely, restored in nuclear medium.
Accordingly hadrons change their properties as demended by the partial restoration. 
Actually experimental data of pionic atoms~\cite{Suzuki:2002ae}  
and pion-nucleus elastic scatterings~\cite{Friedman:2004jh} 
together with theoretical consideration~\cite{Kolomeitsev:2002gc,Jido:2008bk}
have suggested that chiral symmetry is partially restored in nuclear matter with 30\% reduction of the magnitude of
the quark condensate. This is consistent with the prior theoretical studies~\cite{Coh91,Dru91,Bro96,Kai08,Mei02}.

When chiral symmetry is partially restored, one expects the following phenomena in hadron properties:
First of all, the reduction of the mass difference between parity parters, such as 
$\pi$-$\sigma$, $\rho$-$a_{1}$ and $N$-$N^{*}$, are expected, because these parity partners should 
degenerate when chiral symmetry is completely restored 
in the chiral limit. For example, chiral symmetry 
for the nucleon and its parity partners is discussed in Ref.~\cite{Jido:2001nt}.
The $\eta$ mesonic nuclei, which are systems of a $\eta$ meson and a nucleus governed by the strong interaction,
are good to see the possible reduction of the mass difference among the nucleon chiral partners, since
the $\eta$-nucleon systems strongly couples to $N(1535)$, and $N(1535)$ is a candidate 
of the parity partner of nucleon. 
Second of all, one expects a large wave function renormalization of the Nambu-Goldstone boson. 
Since, according to the low energy theorem of chiral symmetry, 
the amplitudes of the Nambu-Goldstone bosons are written in terms of energy expansion in low-energy, 
the in-medium self-energy of the Nambu-Goldstone bosons has strong energy dependence.
This provides a large wave function renormalization. This can be seen in the $K^{+}A$ elastic scattering 
and the $\pi^{0}$ decay in medium. 
Finally, one expects the reduction of the hadron mass generated by chiral symmetry breaking.
For instance, a part of nucleon mass is considered to be generated by spontaneous breaking of chiral symmetry. 
Such a mass should decrease when chiral symmetry is restored. 
Actually a part of the $\eta^{\prime}$ meson is also generated by chiral symmetry breaking~\cite{Jido:2011pq}.
Thus the $\eta^{\prime}$ mass is expected to be reduced in nuclear medium.

\section{Mesons in nuclei}
\subsection{$\bar K$ meson}

There are already a lot of studies on in-medium kaon~\cite{Friedman:2007zz}. 
It is known that the $\bar K N$ interaction is so attractive that the $\Lambda(1405)$ resonance 
can be considered as a quasi-bound state of $\bar KN$~\cite{Hyodo:2011ur}. 
Thus, it is natural to expect that bound states of the $\bar K$ meson and a nucleus exist.  
One of the difficulties in observation of the bound states, even if they exist, is 
that $\bar K$ has large nuclear absorption. 
The one-body absorption into $\pi$ and hyperon, $\bar KN \to \pi Y$, is substantially large, 
and also the nonmesonic two-body absorption, $\bar KNN \to YN$, is not small contribution 
as reported that the two-body absorption should be
30\% of total absorption at saturation density~\cite{Sekihara:2012wj}. 
This strong absorption comes through the $\Lambda(1405)$ resonance.
It may be hard to identify $\bar K$-nucleus bound states owing to the large decay width.
To understand the in-medium properties of $\bar K$,
it is extremely important to pin down the $\bar KN$ interaction as a fundamental interaction 
of the $\bar K$-nucleus system.
For this purpose, the nature of the $\Lambda(1405)$ resonance should be understood very 
well, since  it is sitting at 30 MeV below the threshold of $\bar KN$.
Chiral symmetry determines low-energy $\bar KN$ interaction, but the partial restoration of chiral symmetry 
may play a minor role on the in-medium $\bar K$ properties, 
because dynamics of $\bar KN$, or $\Lambda(1405)$, is much more significant.

The model independent Tomozawa-Weinberg interaction of 
the chiral perturbation theory is attractive enough to form a bound state in the $\bar KN$ 
channel with isospin $I=0$. Theoretically, the $\bar KN$ interaction with $I=0$, however, is not 
so strong that the bound state is formed at 30 MeV below the $\bar KN$ threshold,
which corresponds to 1405 MeV, but the theory predicts the bound state sitting at about 
15 MeV below the threshold, which is 1420 MeV. To obtain the observed $\Lambda(1405)$ spectrum, 
coupled channel analysis of $\bar KN$ and $\pi\Sigma$ is unavoidable, since 
the $\pi\Sigma$ channel is open at the $\Lambda(1405)$ resonance and 
the $\Bar KN$-$\pi\Sigma$ channel coupling is not negligible. 
The coupled-channel analyses based on chiral dynamics~\cite{Kaiser:1995eg} reproduce the observed $\Lambda(1405)$ 
spectrum by the two-pole structure for $\Lambda(1405)$, in which there are two independent states 
with the same quantum number with $\Lambda(1405)$ at around 1420 MeV with a small width
as a bound state of $\bar KN$ and at around 1390 MeV with a large width as 
a $s$-wave resonance of $\pi\Sigma$, and the $\Lambda(1405)$ spectrum is shown up
as an interference of these two state~\cite{Jido:2003cb,Hyodo:2007jq}. 
(This means that the nominal $\Lambda(1405)$ at 1405 MeV with
50 MeV width is not an eigenstate of the Hamiltonian but the two states are.) 
Because the $\bar KN$ scattering state more strongly couples 
to the $\bar KN$ bound state rather than the $\pi\Sigma$ resonance state, 
the double pole scenario for $\Lambda(1405)$ suggests that the peak position corresponding 
to $\Lambda(1405)$ in the spectrum in the $\bar KN$ channel 
appears around 1420 MeV of the bound state energy not at 1405 MeV. 

One of the ways to confirm this scenario is to observe $\Lambda(1405)$ produced by the $\bar KN$ channel.
Nevertheless, because $\Lambda(1405)$ is located below the $\bar KN$ threshold, $\Lambda(1405)$ cannot
be produced kinematically by the direct $\bar KN$ reaction. To create $\Lambda(1405)$ by $\bar KN$,
one makes use of nuclear effects. Several years ago, a $K^-d$ reaction has been proposed~\cite{Jido:2009jf}. 
In this reaction, since strangeness 
is brought from the outside through the $K^{-}$ beam, the $\Lambda(1405)$ resonance is 
produced mainly by $\bar KN$, and the contribution of the $\pi\Sigma$ channel to create $\Lambda(1405)$
is small because multiple steps are necessary to pass the strangeness brought into the system
by $K^{-}$ to the baryon.  
The J-PARC E31 experiment has already started and a preliminary result was reported in this conference~\cite{Inoue}.
The important thing is that, to extract the $\bar KN$ scattering amplitude, one has to make 
the production mechanism of $\Lambda(1405)$ well under control. 
A lot of theoretical progress were reported in this conference~\cite{Ohnishi,Ohnishi:2015iaq,Miyagawa,Miyagawa:2012xz}.

\subsection{$K^{+}$ meson}

It is known that the $K^{+}N$ interaction is repulsive and the $K^{+}N$ cross section is small compared 
to $K^{-}N$~\cite{Dover:1982zh,Friedman:2007zz}. 
The mean free path of the kaon in nuclear medium is estimated to be around 5 fm, 
which is larger than the size of medium-heavy nuclei. 
With these properties, kaon has been considered to be a clean hadronic probe to investigate nuclear matter. 
Here we would like to draw your attention to another characteristic of kaon. 
In contrast to $\bar K$, there are no hyperon resonances strongly coupled to $KN$. 
Thus it is relatively ease to extract in-medium effects on kaon without suffering the strong nuclear absorption. 
This gives us an idea that the $K$ meson is a good probe to investigate (anti-)strangeness in nuclear medium.

Bearing in mind that the mean free path of kaon in nucleus is large, 
we expect that the $K^{+}$-nucleus scattering could be written well by a single 
step $K^{+} N$ interaction  
and multiple scatterings should be strongly suppressed. 
It is surprising, however, that the ratio of the total cross sections 
for $K^{+}$ elastic scattering off carbon and deuterium per nucleon is larger than unity 
in the range of laboratory momenta 450 to 900 MeV/c~\cite{Bugg:1968zz,Weise:1989jq,Weise:1989um}.
The theoretical study in the impulse approximation tells us that
the ratio rather should be suppressed owing to the nuclear shadowing effect. 
This observation implies that the linear density approximation for the $K^{+}$ in nuclear matter 
is broken down in spite of the small interaction between $K^{+}$ and nucleon. 
It is also known that the so-called low density $T\rho$ approximation for the optical potential 
is also broken down~\cite{Friedman:2007zz}, 
which is a consequence of the linear density approximation.
Having extracted the $K^{+}$-nucleus scattering amplitude 
by fitting the $K^{+}A$ elastic scattering data with the optical potential in the $T\rho$ approximation,
Ref.~\cite{Friedman:2007zz} has concluded that the $K^{+}$-nucleus scattering amplitude 
is repulsively enhanced in about 15\% than the $K^{+}N$ scattering amplitude.
This is also seen in pionic atoms as ``missing repulsion'' that 
the isospin-odd $\pi^{-}$-nucleus scattering length is unexpectedly more repulsive 
than the $\pi^{-}N$ scattering length. 
Several possible explanations for the repulsive enhancement have been proposed: 
For instance, this is due to nucleon-nucleon correlation, ``swelling'' of nucleon~\cite{Siegel:1985hp}
and mass reduction of vector mesons caused 
by the scale change in a nuclear medium~\cite{Brown:1988gu}. 

Here we would like to explain this enhancement as the wave function renormalization~\cite{Aoki}. 
If the in-medium self-energy has strong energy-dependence, the wave function renormalization 
plays an important role for the in-medium effect as one of the next-to-leading corrections of the
linear density approximation~\cite{Kolomeitsev:2002gc}. Especially, for the Nambu-Goldstone
bosons, the interactions should vanish in the soft limit ($p_{\mu} \to 0$) and the chiral limit 
(Adler zero), and thus the self-energy for the Nambu-Goldstone boson should have 
substantial energy dependence~\cite{Jido:2000bw,Jido:2008bk}. 
According to the argument by Kolomeitsev developed for pion~\cite{Kolomeitsev:2002gc},
we obtain the energy-independent optical potential for the in-medium kaon as follows.
The in-medium dispersion relation for $K^{+}$ at rest  
determines the in-medium $K^{+}$ mass as the solution of $m^{*2} - m^2 - \Sigma(m^{*}) = 0$.
The optical potential for the in-medium kaon is given by the self-energy 
at the kaon energy $\omega=m^{*}$ as $2m V_{\rm opt}(m^{*}) = \Sigma(m^{*})$.
Now assuming $V_{\rm opt} \ll m$, that is, $m^{*} \approx m$, 
we expand the self-energy around $\omega = m$, set $\omega = m^{*}$ and obtain
\begin{equation}
2 m V_{\rm opt}(m^*) = \Sigma(m) + (m^{*2} - m^2) \left.\frac{\partial \Sigma}{\partial \omega^2} \right|_{\omega=m}+ \cdots 
\simeq \left(1 + \frac{\partial \Sigma}{\partial \omega^2}\right) \Sigma(m) 
\simeq Z \Sigma(m),
\end{equation}
where in the third equation we have neglected the higher order of $m^{*2} - m^2$
and assumed that the difference of $\Sigma(m^{*})$ and $\Sigma(m)$ gives a higher 
order contribution of the density expansion, 
and in the fourth equation we have introduced the wave function renormalization
\begin{equation}
   Z = \left. \left( 1 - \frac{\partial \Sigma}{\partial \omega^2}\right)^{-1} \right|_{\omega= m}.
\end{equation}
In this way, when the energy dependence of the self-energy is strong, which is the case for 
the Nambu-Goldstone boson, the wave function renormalization becomes one of the 
essential medium effects beyond the linear density approximation. 
Finally the optical potential is given by 
\begin{equation}
2 m V_{\rm opt} =  Z \rho T_{K^+ N} .
\end{equation}

Let us now evaluate the $K^{+}N$ scattering amplitude in the chiral perturbation theory.
The leading order contribution is calculated by the Tomozawa-Weinberg interaction as
\begin{equation}
   T^{I=0}_{KN} = 0, \qquad  T^{I=1}_{KN} = \frac{\kslash + \kslash^{\prime}}{2 f_{K}^{2}},
\end{equation}
for the isospin 0 and 1 $KN$ channels, respectively. The averaged $K^{+}N$ scattering 
amplitude is given by 
\begin{equation}
 T_{K^{+}N} = \frac{1}{2} ( T_{K^{+}p} + T_{K^{+}n} ) 
 = \frac{1}{4} (3T^{I=1}_{KN}+ T^{I=0}_{KN}),
\end{equation}
and thus around the threshold the forward $K^{+}N$ scattering amplitude is written as 
\begin{equation}
   T_{K^{+}N}(\omega) = \frac{3}{4} \frac{\omega}{f_{K}^{2}}.
\end{equation}
With this amplitude, using the linear density approximation for the kaon self-energy 
in symmetric nuclear matter with density $\rho$,
$\Sigma(\omega) = T_{K^{+}N}(\omega) \rho$, we obtain
\begin{equation}
   Z = \left( 1 -  \rho \left.\frac{\partial T_{K^{+}N}}{\partial \omega^{2}}\right|_{\omega = m_{K^{+}}} \right)^{-1}
   =  \left( 1 - \frac{1}{2m_{K^{+}}} \frac{3}{4 f_{K}^{2}} \rho  \right)^{-1}
   \approx 1 + 0.082 \frac{\rho}{\rho_{c}}
\end{equation}
where we have used $m_{K}=493.7$ MeV, $f_{K} = 110$ MeV and $\rho_{c} = 0.17$ fm$^{-3}$.
There around 10\% enhancement of the optical potential is explained by the wave function 
renormalization.

The in-medium correction from the wave function renormalization is more significant in 
the $\pi^{0}$ decay~\cite{Goda:2013npa,Jenifer}.  
The in-medium amplitude of the  $\pi^{0}$ decay into $\gamma\gamma$ is written 
as the one-particle irreducible vertex correction and the wave function renormalization,
$M^*_{\gamma\gamma} = \sqrt Z \hat M_{\gamma \gamma}$~\cite{Goda:2013npa}. 
It is known that there is no vertex correction in linear density approximation~\cite{Meissner:2001gz},
and thus the in-medium change of the amplitude is given solely by the wave function 
renormalization, $M^*_{\gamma\gamma} = \sqrt Z M_{\gamma \gamma}$. If one neglects 
the change of the phase space of the $\pi^{0}$ decay in the medium and uses the 
result of the in-medium wave function renormalization obtained in Ref.~\cite{Goda:2013npa}, 
one finds 
\begin{equation}
  \frac{\Gamma^*_{\gamma\gamma}}{\Gamma_{\gamma \gamma}} = Z \simeq 1 + 0.4 \frac{\rho}{\rho_0},
\end{equation}
and one expects the 40\% enhancement of the $\pi^{0}$ decay into $\gamma\gamma$ in  
nuclear matter.

\subsection{$\eta$ meson}

\begin{figure}[t]
\begin{center}
\includegraphics[width=0.6\textwidth]{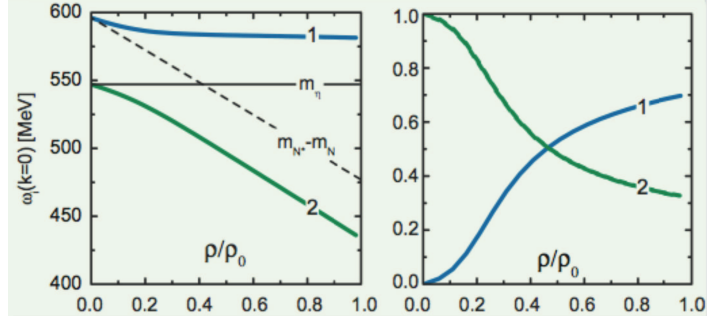}
\end{center}
\caption{
(left) The real part of the pole positions of the in-medium $\eta$ propagator as a function 
of the density $\rho$ in units of the saturation density $\rho_{0}$. The thick solid lines with label 1 and 2
correspond to the modes appearing in the nuclear medium as a consequence of the coupling between 
the $\eta$ and $N^{*}$-$h$ modes. 
The thin solid and dashed lines denote the density dependence of the $\eta$ and $N^{*}$-$h$ modes, respectively.
(right) The residue of the in-medium $\eta$ propagator at the pole position of each mode. The residue of the propagator 
expresses the height of the peak for each mode in the spectral function.
\label{fig:levelcross}}
\end{figure}

The $\eta$ mesonic nuclei, which are bound systems of $\eta$ in nuclei, were firstly predicted by Haider 
and Liu~\cite{Haider:1986sa}. The hadron-nucleus bound systems are good ``laboratories'' to investigate 
hadrons in nuclei. The $\eta$ meson is also one of the Nambu-Goldstone bosons associated with 
the spontaneous breaking of the SU(3) chiral symmetry. The Tomozawa-Weinberg interaction 
of the $\eta N$ channel is, however, null, and thus the contribution from the leading-order 
term of the chiral perturbation theory vanishes. Instead, the $\eta N$ system has a strong 
coupling to the $N^{*}(1535)$ resonance.
The excitation energy of $\eta N \to N^{*}(1535)$ is so small that it is just about 50 MeV 
when we evaluate it in vacuum, in contrast to the $\Delta$ resonance, which is located 
150 MeV above $\pi N$. 
Thus, thanks to the small excitation energy, for the in-medium $\eta$ meson, the channel coupling 
to the $N^{*}(1535)$--$N$-hole ($N^{*}$-$h$) mode should be unavoidably taken into account. 
In the context of chiral symmetry, 
the $N^{*}(1535)$ resonance can be considered a chiral partner 
of the nucleon~\cite{DeTar:1988kn,Jido:2001nt,Kim:1998up}. 
This implies that we can learn also the in-medium properties of $N^{*}(1535)$
and investigate the chiral symmetry for the nucleons by studying the $\eta$ meson 
in the nuclear medium where chiral symmetry is partially restored~\cite{Jido:2002yb}.

\begin{figure}[t]
\begin{center}
\includegraphics[width=0.9\textwidth]{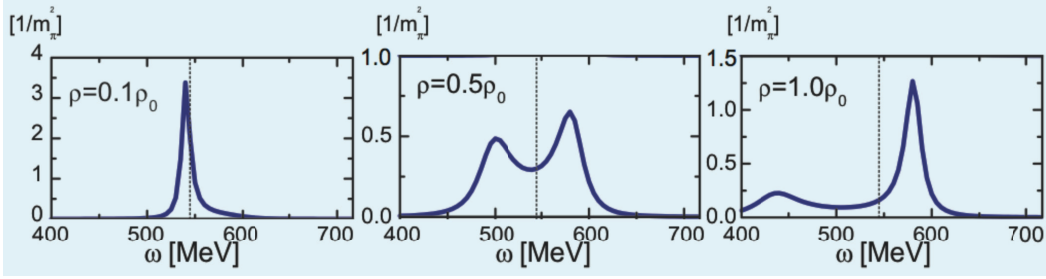}
\end{center}
\caption{ The spectral function $S_{\eta}(\omega, \rho) = - {\rm Im} G_{\eta}(\omega, \rho)$ with $k=0$ calculated
by the $N^{*}$ dominance model with the chiral doublet picture for $N$ and $N^{*}$ for the densities $\rho = 0.1\rho_{0}$, $0.5 \rho_{0}$ and $\rho_{0}$.\label{fig:spec}}
\end{figure}

If the $N^*(1535)$ resonance is the chiral partner of nucleon, the mass difference of $N$ and $N^*$ 
should be reduced as chiral symmetry is partially restored in the nuclear matter. The reduction is 
estimated as about 150 MeV at the saturation density in a chiral doublet model~\cite{Kim:1998up}. 
This implies that the energy of the $N^{*}$-$h$ mode gets lower in nuclear matter and
at a certain density the level crossing of the $\eta$ and $N^{*}$-$h$ modes takes place~\cite{Jido:2008ng}
as shown as the thin solid and dashed lines in Fig.~\ref{fig:levelcross}. 
These two modes couple each other in the nuclear medium. The mode energies are calculated as a pole 
position of the in-medium $\eta$ propagator
 $ G_{\eta} (\omega, k ; \rho) = i/({\omega^{2} - k^{2} - m_{\eta}^{2} - \Sigma_{\eta}(\rho)} )$
with the in-medium self-energy $\Sigma_{\eta}(\rho)$.  Here we take a $N^{*}(1535)$ dominance model 
for the self-energy $\Sigma_{\eta}$ as
\begin{equation}
   \Sigma_{\eta}(\omega, \rho) = g_{\eta}^{2} \frac{\rho}{\omega + m_{N}^{*}(\rho) - m_{N^{*}}^{*}(\rho) 
   + i\Gamma_{N^{*}}(\omega, \rho)/2}
\end{equation}
where $g_{\eta}$ is the coupling strength of the $s$-wave $\eta NN^{*}(1535)$ coupling,
$m_{N}^{*}$ and $m_{N^{*}}^{*}$ denote the in-medium masses of nucleon and $N^{*}(1535)$, respectively,
and the $\Gamma_{N^{*}}$ is the $N^{*}(1535)$ decay width, which depends on the energy and density. 
The real parts of the pole positions are plotted as the thick solid lines 
in the left panel of Fig.~\ref{fig:levelcross}. These two modes are labeled 1 and 2 in Fig.~\ref{fig:levelcross}.
With the mode coupling, one can see level repulsion in which 
one of the modes is pushed above repulsively in higher densities. 
It is also interesting that, as a consequence of 
the level crossing, the strength of these modes changes drastically  as the density increases. 
In the right panel of Fig.~\ref{fig:levelcross}, we plot 
the residua of the propagator $G_{\eta}$ at the pole positions, which are calculated by
\begin{equation}
   Z = \left( 1 - \frac{\partial \Sigma_{\eta}}{\partial \omega^{2}} \right)^{-1}.
\end{equation}
The residue of the propagator corresponds to the renormalization of the wave function of each mode.
In vacuum, the wave function for the $\eta$ meson, or the residue of the $\eta$ mode, is normalized 
as unity, while the $N^{*}$-$h$ mode is absence in vacuum and its wave function is zero. 
Owing to the mode coupling, the wave functions of the two modes mix each other
and the wave functions are renormalized.
The mode originating from the $\eta$ meson in vacuum, labeled 2 in Fig.~\ref{fig:levelcross}, 
has a larger residue in lower densities and the residue is reduced as the density increases, 
while the other mode labeled 1 in Fig.~\ref{fig:levelcross}, 
which originates from the $N^{*}$-$h$ mode at $\rho=0$, 
has a smaller residue in lower densities and the residue gets enhanced in higher densities.
Thus, for the in-medium $\eta$ meson, which is an admixture of the $\eta$ and $N^{*}$-$h$ modes, 
looks attractive in lower densities and turns repulsive in higher densities. 
The spectral function $S_{\eta}(\omega, \rho) = - {\rm Im} G_{\eta}(\omega, \rho)$ with $k=0$ is 
plotted in Fig.~\ref{fig:spec}.
The level crossing is expected to be observed in nuclear reactions, such as $(\gamma, p)$, $(d, ^{3}{\rm He})$
and $(\pi, N)$ with nuclear targets~\cite{Jido:2008ng,Hayano:1998sy,Nagahiro:2003iv,Nagahiro:2005gf,Nagahiro:2008rj}.

As another picture of $N^{*}(1535)$, it could be a dynamically generated resonance 
of the octet mesons and baryons. In such a case, it is reported that 
the $N^{*}(1535)$ resonance has large components of $K\Lambda$ and $K\Sigma$~\cite{Inoue:2001ip} and
is insensitive to the medium effects since the $\Lambda$ and $\Sigma$ are free from the Pauli blocking
by nucleons, which is one of the main medium effects~\cite{Inoue:2002xw}. In this case the level crossing 
does not take place. 
These two pictures are distinguishable when one observes the missing mass spectroscopy of 
the formation reactions of the $\eta$ mesonic nuclei~\cite{Jido:2008ng,Nagahiro:2008rj}. 
One of the main differences can be seen in the formation spectrum around the quasi-free $\eta$ energy region. 
When the level crossing takes place, the $\eta$ mode feels repulsive interaction since the $N^{*}$-$h$ mode
comes down below the $\eta$ mode. Accordingly some enhancement appears above the $\eta$ creation 
threshold.  When the level crossing does not take place, such enhancement does not show up.

\subsection{$\eta^{\prime}$ meson}

\begin{figure}[t]
\begin{center}
\includegraphics[width=0.6\textwidth]{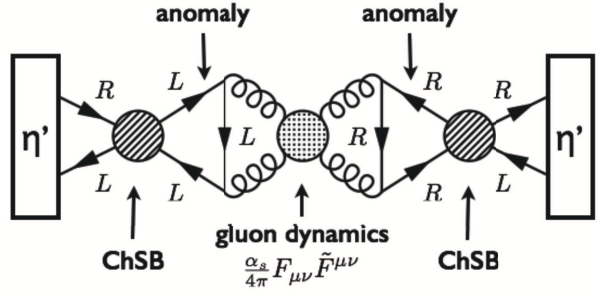}
\end{center}
\caption{Schematic figure of the two-point function of the $\eta^{\prime}$ field. 
To couple to the axial U(1) breaking effect caused by the quantum anomaly of the gluonic field,
chiral symmetry is necessarily broken. \label{fig:etap}}
\end{figure}

\begin{figure}
\begin{center}
\includegraphics[width=0.5\textwidth]{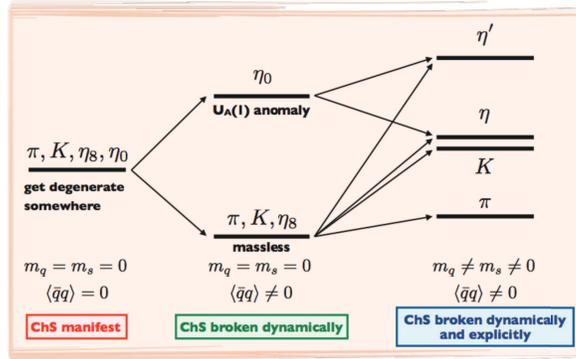}
\end{center}
\caption{The fate of the SU(3) nonet pseudoscalar meson masses in different breaking patterns of 
chiral symmetry. \label{fig:nonetmass}}
\end{figure}

The $\eta^{\prime}$ meson would be one of the Nambu-Goldstone bosons 
associated with spontaneous breaking of the three flavor chiral symmetry 
in classical theory of chromodynamics. In fact, quantum chromodynamics
has no axial U(1) symmetry at the beginning and it is explicitly broken by
quantum effect. Thus, the $\eta^{\prime}$ meson does not have to be a
Nambu-Goldstone boson of the chiral symmetry breaking and 
is not massless even in the chiral limit. The observed $\eta^{\prime}$ meson
mass is 958 MeV/c$^{2}$ as large as the proton mass. 
Actually the $\eta^{\prime}$ mass has a strong connection also 
to the breaking of the SU(3) chiral symmetry~\cite{Lee:1996zy,Jido:2011pq}.
Since the $\eta^{\prime}$ meson is a pseudoscalar boson and the $\eta'$ field is
written in terms of both the left and right handed quark fields,
it cannot couple to nonchiral gluonic field, which causes the axial U(1) anomaly, 
without breaking chiral symmetry in any sense, explicitly and/or spontaneously.  
Therefore, to generate the $\eta^{\prime}$ mass, the axial U(1) anomaly is not 
the only source, but also the SU(3) chiral symmetry is necessarily broken,
as schematically shown in Fig.~\ref{fig:etap}. This means that the 
mass gap of the $\eta$ and $\eta^{\prime}$ is generated by the chiral 
symmetry breaking through the axial U(1) anomaly. 
Therefore, when the broken chiral symmetry is restored in the case of the 
chiral limit, the $\eta^{\prime}$ meson should get degenerate with the 
octet pseudoscalar mesons as shown in Fig.~\ref{fig:nonetmass}.

This scenario suggests that the mass gap of $\eta$ and $\eta^{\prime}$
gets reduced in the nuclear medium where the spontaneously broken chiral symmetry 
is to be partially restored. Assuming that the $\eta$ meson mass is insensitive to the 
partial restoration of chiral symmetry because the $\eta$ mass is generated by the explicit 
chiral symmetry breaking through the quark mass, we expect that the $\eta^{\prime}$ mass
gets reduced in the nuclear medium. According to a simple estimation done in Ref.~\cite{Sakai:2013nba},
100 MeV reduction of the $\eta^{\prime}$ mass is expected at the nuclear saturation density, 
where one expects 35\% reduction of the magnitude of the quark condensate. 
The linear sigma model analysis also suggests 80 MeV reduction~\cite{Sakai:2013nba}, and
the NJL models predict 150 MeV reduction~\cite{Costa:2002gk,Nagahiro:2006dr}.
With such enough attraction for the $\eta^{\prime}$ meson in the nuclear medium,
one expects $\eta^{\prime}$-nucleus bound systems. The first calculation of the formation spectra
was done in Ref.~\cite{Nagahiro:2004qz} for a $(\gamma, p)$ reaction with $^{12}$C targets.
Later $(p,d)$ reactions have been studied~\cite{Nagahiro:2012aq} and
recently, the feasibility study of observing $\eta^{\prime}$ mesonic nuclei has been done in Ref.~\cite{Itahashi:2012ut}.

In the view of the linear sigma model~\cite{Sakai:2013nba}, (at least a part of) the nucleon mass is generated by
the spontaneous breaking of chiral symmetry where the chiral filed $\sigma$ gets condensed 
in vacuum, and the nucleon is expressed by $m_{N} = - g\langle \sigma_{0} \rangle$. 
This implies the presence of the strong $\sigma NN$ coupling. This is the origin of the 
scalar attraction in the $NN$ interaction. In the same way, the chiral symmetry breaking
generates a part of the $\eta^{\prime}$ mass with the help of the axial anomaly. 
This implies again that there exists a strong coupling of $\sigma \eta^{\prime} \eta^{\prime}$. 
This leads to a strong attraction in the $\eta^{\prime}$-$N$ interaction in the scalar
channel by exchanging the $\sigma$ meson. It is also known that the Tomozawa-Weinberg
interaction, which is the leading term of the chiral perturbation 
theory, is zero for the $\eta^{\prime}N$ channel. Thus, there may be no (or very small) repulsion 
in the vector channel, which plays an important role for the $NN$ interaction. 
With this knowledge, a two body bound state of $\eta^{\prime}N$ is evaluated 
in the linear sigma model using the same machinery in which $\Lambda(1405)$ is calculated
as a quasi-bound state of $\bar KN$, and it is found that two-body $\eta^{\prime}N$ bound state
is obtained with 6 MeV binding energy~\cite{Sakai:2013nba} and 
that with the coupled channel effect of $\eta^{\prime}N$ and 
$\eta N$ the bound state is found with 12 MeV binding energy and 6 MeV width~\cite{Sakai:2014zoa}.

\section{Summary}

We have discussed the pseudoscalar mesons in the nuclear medium under the situation that
chiral symmetry is partially (30\%) restored. Such in-medium changes of the meson properties
are observed in meson and nucleus systems.
Our expectations of the physical consequences when the partial restoration of 
chiral symmetry takes place in the nuclear medium are as follows:

{\bf Reduction of the mass difference among the chiral partners:}
The chiral symmetry breaking resolves parity degeneracy in the hadron mass spectrum
and thus, it is responsible for the mass splitting of the chiral partners. When the broken chiral symmetry 
is restored in the nuclear medium, the mass gap should get smaller than that in vacuum. 
The $N^{*}(1535)$ nucleon resonance can be regarded as the chiral partner of nucleon,
and the mass difference of $N^{*}(1535)$ and $N$ gets reduced in this case. 
The reduction of the mass gap could be observed in the formation spectrum of 
the $\eta$ mesonic nuclei, since the $\eta$ meson couples to the $N^{*}$-$h$ mode in the
nuclear medium.

{\bf Substantial effect  from the wave function renormalization of the Nambu-Goldstone bosons:}
The low energy theorem of chiral symmetry tells us that the interactions of the Nambu-Goldstone 
bosons are described in terms of energy expansion. Thus, the in-medium self-energy 
of the Nambu-Goldstone boson necessarily has energy dependence, and 
the wave function renormalization can be a substantial medium effect as one of the corrections
beyond the linear density approximation. We have seen that the enhancement
of the $K^{+}$-nucleus scattering can be partially explained by the $K^{+}$ wave function 
renormalization. We have also seen that one expects a large enhancement of $\pi^{0} \to \gamma\gamma$
in nuclear medium due to the large wave function renormalization for $\pi^{0}$.

{\bf Reduction of hadron mass:}
The masses of some hadrons are generated by the spontaneous breaking of chiral symmetry, such as nucleon. 
A part of the $\eta^{\prime}$ mass is generated by the SU(3) chiral symmetry breaking through the axial U(1) anomaly. 
A 100 MeV reduction of the $\eta^{\prime}$ mass at the saturation density is expected, 
and this provides an attractive optical potential for $\eta^{\prime}$ meson and nucleus systems.
Thus, $\eta^{\prime}$ bound states are expected with this attractive potential. 
In addition, the mass generation by the spontaneous chiral symmetry breaking implies 
a strong coupling to the $\sigma$ field. This leads to strong attraction of the $\eta^{\prime}$-N interaction
from the isoscalar-scalar $\sigma$ exchange. With this attraction, a two-body $\eta^{\prime}N$
bound state can be formed with a several MeV binding energy and a few MeV width.

\section*{Ackonowlegdement}
The author would like to express his cordial appreciation to his collaborators, 
T. Hyodo, E. Oset, T. Sekihara, J.A. Oller, A. Ramos, U.G. Meissner, 
K. Aoki, S. Goda, J. Nebreda, 
S. Hirenzaki, H. Nagahiro, E.E. Kolomeitsev, K. Itahashi, H. Fujioka,
S. Sakai.
This work and the presence at this conference 
were financially supported by Grants-in-Aid for Scientific Research (No. 25400254).


\begin{thebibliography}{99}

\bibitem{Suzuki:2002ae}
K. Suzuki {\it et al.}:
Phys. Rev. Lett. {\bf 92} (2004) 072302.


\bibitem{Friedman:2004jh}
E. Friedman {\it et al.}:
Phys. Rev. Lett. {\bf 93} (2004) 122302.

\bibitem{Kolomeitsev:2002gc}
E.E. Kolomeitsev, N. Kaiser, and W. Weise:
Phys. Rev. Lett. {\bf 90} (2003) 092501.

\bibitem{Jido:2008bk}
D. Jido, T. Hatsuda, and T. Kunihiro:
Phys. Lett. {\bf B670} (2008) 109-113.

\bibitem{Coh91}
T.D. Cohen, R.J. Furnstahl, and D.K. Griegel: Phys. Rev. C {\bf 45} (1992) 1881.

\bibitem{Dru91}
E.G. Drukarev, and E.M. Levin: Prog. Part. Nucl. Phys. {\bf 27} (1991) 77.

\bibitem{Bro96}
R. Brockmann, and W. Weise: Phys. Lett. B {\bf 367} (1996) 40.

\bibitem{Kai08}
N. Kaiser, P. de Homont, and W. Weise: Phys. Rev. C {\bf 77} (2008) 025204.

\bibitem{Mei02}
U.G. Meissner, J.A. Oller, and A. Wirzba: Annals Phys. {\bf 297} (2002) 27.

\bibitem{Jido:2001nt}
D. Jido, M. Oka, and A. Hosaka,
Prog. Theor. Phys. {\bf 106} (2001) 873.

\bibitem{Jido:2011pq}
D. Jido, H. Nagahiro, and S. Hirenzaki:
Phys. Rev. C {\bf 85} (2012) 032201(R).

\bibitem{Friedman:2007zz}
As a recent review, 
E. Friedman, and A. Gal:
Phys. Rept. {\bf 452} (2007) 89.

\bibitem{Hyodo:2011ur}
As a recent review, T. Hyodo and D. Jido:
Prog. Part. Nucl. Phys. {\bf 67} (2012) 55.

\bibitem{Sekihara:2012wj}
T. Sekihara, J. Yamagata-Sekihara, D. Jido, and Y. Kanada-En'yo:
Phys. Rev. C {\bf 86} (2012) 065205.

\bibitem{Kaiser:1995eg}
N.~Kaiser, P. B. Siegel, and W. Weise:
Nucl. Phys. {\bf A594} (1995) 325-345.


\bibitem{Jido:2003cb}
D. Jido, J. A. Oller, E. Oset, A. Ramos, and U. G. Meissner:
Nucl. Phys. {\bf A725} (2003) 181.

\bibitem{Hyodo:2007jq}
T. Hyodo and W. Weise:
Phys. Rev. {\bf C77} (2008) 035204.

\bibitem{Jido:2009jf}
D. Jido, E. Oset, and T. Sekihara:
Eur. Phys. J. {\bf A42} (2009) 257.


\bibitem{Inoue}
K. Inoue: in this proceedings.

\bibitem{Ohnishi}
S. Ohnishi: in this proceedings.

\bibitem{Ohnishi:2015iaq}
  S.~Ohnishi, Y.~Ikeda, T.~Hyodo and W.~Weise:
  arXiv:1512.00123 [nucl-th].

\bibitem{Miyagawa}
K. Miyagawa: in this proceedings.

\bibitem{Miyagawa:2012xz}
  K.~Miyagawa and J.~Haidenbauer:
  Phys.\ Rev.\ C {\bf 85} (2012) 065201.





\bibitem{Dover:1982zh}
C.B. Dover, and G.E. Walker:
Phys. Rept. {\bf 89} (1982) 1.

\bibitem{Bugg:1968zz}
D.V. Bugg, {\it et al.}:
Phys. Rev. {\bf 168} (1968) 1466.

\bibitem{Weise:1989jq}
  W.~Weise: Proceedings of IUPAP International Nuclear Physics Conference,
  Sao Paulo, Brazil, 1989, Vol.~2, 211.

\bibitem{Weise:1989um}
W. Weise:
Nuovo Cim. {\bf A102} (1989) 265.

\bibitem{Siegel:1985hp}
P.B. Siegel, W.B. Kaufmann, and W.R. Gibbs:
Phys. Rev. {\bf C31} (1985) 2184.

\bibitem{Brown:1988gu}
G.E. Brown, C.B. Dover, P.B. Siegel, and W. Weise:
Phys. Rev. Lett. {\bf 60} (1988) 2723.

\bibitem{Aoki}
K. Aoki, and D. Jido, in preparation.

\bibitem{Jido:2000bw}
D. Jido, T. Hatsuda, and T. Kunihiro:
Phys. Rev. {\bf D63} (2001) 011901.

\bibitem{Goda:2013npa}
S. Goda, and D. Jido:
Prog. Theor. Exp. Phys. {\bf 2014} (2014) 033D03.

\bibitem{Jenifer}
J. Nebreda, and D. Jido: in preparation.

\bibitem{Meissner:2001gz}
U.G. Meissner, J.A. Oller, and A. Wirzba:
Annals Phys. {\bf 297} (2002) 27.

\bibitem{Haider:1986sa}
Q. Haider, and L. C. Liu:
Phys. Lett. {\bf B172} (1986) 257.

\bibitem{DeTar:1988kn}
C.E. DeTar and T. Kunihiro:
Phys. Rev. {\bf D39} (1989) 2805.

\bibitem{Kim:1998up}
Hung-chong Kim, D. Jido, and M. Oka:
Nucl. Phys. {\bf A640} (1998) 77.

\bibitem{Jido:2002yb}
D. Jido, H. Nagahiro, and S. Hirenzaki:
Phys. Rev. {\bf C66} (2002) 045202.


\bibitem{Jido:2008ng}
D. Jido, E.E. Kolomeitsev, H. Nagahiro, and S. Hirenzaki:
Nucl. Phys. {\bf A811} (2008) 158.

\bibitem{Hayano:1998sy}
R. S. Hayano, S. Hirenzaki, and A. Gillitzer:
Eur. Phys. J. {\bf A6} (1999) 99.

\bibitem{Nagahiro:2003iv}
H. Nagahiro, D. Jido, and S. Hirenzaki:
Phys. Rev. {\bf C68} (2003) 035205.


\bibitem{Nagahiro:2005gf}
H. Nagahiro, D. Jido, and S. Hirenzaki:
Nucl. Phys. {\bf A761} (2005) 92.

\bibitem{Nagahiro:2008rj}
H. Nagahiro, D. Jido, and S. Hirenzaki:
Phys.Rev. {\bf C80} (2009) 025205.


\bibitem{Inoue:2001ip}
T. Inoue, E. Oset, and M. J. Vicente Vacas:
Phys. Rev. {\bf C65} (2002) 035204.

\bibitem{Inoue:2002xw}
T. Inoue and E. Oset:
Nucl. Phys. {\bf A710} (2002) 354.




\bibitem{Lee:1996zy}
S.H. Lee, and T. Hatsuda:
Phys. Rev. {\bf D54} (1996) 1871.


\bibitem{Sakai:2013nba}
S. Sakai, and D. Jido:
Phys. Rev. C {\bf 88} (2013) 064906.

\bibitem{Costa:2002gk}
Pedro Costa, Maria C. Ruivo, and Yu. L. Kalinovsky:
Phys. Lett. {\bf B560} (2003) 171.


\bibitem{Nagahiro:2006dr}
H. Nagahiro, M. Takizawa, and S. Hirenzaki:
Phys. Rev. {\bf C74} (2006) 045203.


\bibitem{Nagahiro:2012aq}
H. Nagahiro, D. Jido, H. Fujioka, K. Itahashi, and S. Hirenzaki:
Phys. Rev. C {\bf 87} (2013) 045201.

\bibitem{Nagahiro:2004qz}
H. Nagahiro and S. Hirenzaki:
Phys. Rev. Lett. {\bf 94} (2005) 232503.

\bibitem{Itahashi:2012ut}
K. Itahashi {\it et al.}: 
Prog. Theor. Phys. {\bf 128} (2012) 601.

\bibitem{Sakai:2014zoa}
S. Sakai, and D. Jido:
Hyperfine Interact. {\bf 234} (2015) 71.

%
\end{thebibliography}
\end{document}